\documentclass[twocolumn,showpacs,preprintnumbers,amsmath,amssymb]{revtex4}
\usepackage{graphicx}
\usepackage{dcolumn}

\usepackage{bm}
\begin{document}
\title{Another Origin of Bimodal Size Distribution in InAs Self-Assembled Quantum Dots}
\author{Bhavtosh Bansal$^*$ , M. R. Gokhale, A. Bhattacharya and B. M. Arora}
\affiliation{Department of Condensed Matter Physics and Materials
Science, Tata Institute of Fundamental Research, 1 Homi Bhabha Road, Mumbai-400005,
India}
{
\date{\today}
\begin{abstract}
The evolution of InAs quantum dots grown on InP substrates by metal-organic vapour phase 
epitaxy is studied as a function of InAs coverage. Under specific growth conditions, the 
onset of the two- to three-dimensional transition is seen to proceed via two distinct 
pathways: through (i) an abrupt appearance of quantum dots as expected in the usual 
Stranski-Krastanov growth picture and (ii) a continuous evolution of small surface 
features into well developed quantum dots.  The average size of the features in both these 
families increases with coverage, leading to a bimodal distribution in dot sizes at an 
intermediate stage of growth evolution that eventually becomes a unimodal distribution as 
more material is deposited. Complementary information obtained from independent 
measurements of photoluminescence spectra and surface morphology is correlated and is 
found to be independently consistent with the picture of growth proposed.  
\end{abstract}
\maketitle
The semiconductor heterostructures grown via the strain-mediated Stranski-Krastanov (S-K) 
route\cite{stangl, joyce review} have been of special interest. The resulting self 
assembled clusters are extremely small ($\sim$10nm, hence called quantum dots), coherently 
strained (hence optically active) and have a size dispersion that may be acceptable for 
their use in real optoelectronic devices, e.g., \cite{laser refs, detectors}. The S-K 
transition has been systematically studied in most III-V semiconductors of interest (e.g. 
InAs/GaAs \cite{petroff_prb, {Ramachandran}, {patella}, da silva, duarte}, 
In$_{1-x}$Ga$_{x}$As/GaAs\cite{leon-fafard, cullis}, InAs/InP\cite{ponchet}, 
GaSb/GaAs\cite{muller-kirsch}). While the growth route may be rationalized by energetic 
considerations that predict an initial two-dimensional growth followed by an abrupt 
appearance of three-dimensional clusters upon the `wetting layer'\cite{stangl}, the actual 
process of self-assembly is complicated by kinetic effects, substrate conditions and 
orientation, a non-quiescent wetting layer during growth evolution, alloying and 
long-ranged substrate-mediated elastic interactions. For example, it has also been 
observed, though not consistently by different groups, that there may be redistribution of 
matter leading to a decrease in the quantum dot (QD) size with coverage. Often a bimodal 
size distribution \cite{petroff_prb, ponchet} has also been observed at an intermediate 
growth stage. Although many of these phenomena have been described in terms of elastic 
energy barriers corresponding to change in shape and a maximum permissible size for 
quantum dots beyond which they are dislocated, the complex growth evolution of 
self-assembled dots, especially under far from equilibrium (usual) growth conditions,  is 
still very much an open problem despite more than a decade of extensive work. 

In this paper, we focus on InAs/InP self-assembled QDs grown by metal-organic vapour phase 
epitaxy (MOVPE)\cite{movpe refs1, movpe refs2}. The samples are studied at different 
stages of evolution by independently studying and quantitatively correlating the 
morphology and the optical emission spectra. Contrary to many of previous reports, we 
have, on samples grown under more non-equilibrium conditions (lower temperature, 
450$^\circ$C and higher growth rate $\sim$1-2 monolayers(ML) per second) where surface 
kinetics may be an important limiting step, observed that the 2D-3D transition is not 
completely abrupt and proceeds via two different pathways. This study also yields a 
different picture for bimodality in the QD's size distribution.

Growth was carried out using low pressure (100 torr) MOVPE on n+ doped (001) ``epi-ready" 
InP substrates in a horizontal reactor with hydrogen as the carrier gas. Group III and V 
sources were trimethyl-indium and arsine and phosphine respectively. Prior to InAs growth, 
a InP buffer layer was grown first at 625$^\circ$C and then with the temperature 
continuously ramped down and stabilized to 450$^\circ$C. InAs layers were grown at a 
relatively low temperature of 450$^\circ$C at a growth rate\cite{footnote growth rate} of 
approximately 1.8ML/s. A pair of samples was grown with identically deposited InAs layer 
in two growth runs. In the first case, the sample was taken out of the reactor after InAs 
deposition itself to enable a study of surface morphology and in the second case a InP cap 
layer was grown for samples used for photoluminescence study. For these samples, about 50 
\AA$\,$  InP was deposited at the InAs deposition temperature to avoid any further 
ripening during the higher temperature overgrowth. Surface features were studied ex-situ 
using an atomic force microscope (AFM) in contact mode, typically within a few hours of 
sample growth.  The PL spectra were measured at $\sim$25 K at low enough excitation power 
($\sim 0.5 W/cm^2$) to preclude any subband filling and were corrected for the system 
response using a standard black body source. 
\begin{figure}[!h]
\begin{center}
\resizebox{!}{10cm} {\includegraphics{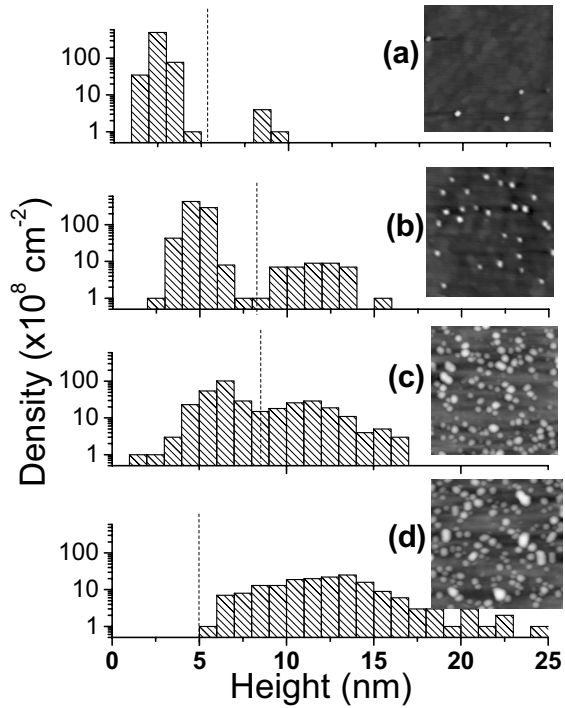}}
\caption{\label{fig:fig2}{\it Histogram of island heights inferred from $1\mu m \times 
1\mu m$ AFM images on samples with increasing InAs coverage. The coverage is 
approximately: (a) 4.5ML, (b) 6.5ML (c)9ML (d)14.5ML. The coverage is calibrated against 
the WL PL peak in Fig.\ref{fig:fig3}(a). See text. The dotted line depicts the separation 
of the islands into two families, evidently evolved from two different routes (see Fig.3). 
This leads to a bimodal size distribution (a)-(c) which merges into a single broad 
distribution (d).}}     
\end{center}
\end{figure}

Fig. \ref{fig:fig2} (inset) shows $1\mu m \times 1\mu m$ AFM scans for samples with 
progressively increasing InAs coverage in Fig. \ref{fig:fig2}(a) to (d). The histograms of 
the island heights corresponding to these AFM images are shown in the main 
Fig.\ref{fig:fig2}. All well-separated ($\sim$20nm) convex features with heights above two 
ML were counted here. Fig.\ref{fig:fig2} (a) depicts an early stage of growth. Here we 
observe that along with very few ($\sim 5\times 10^8$ cm$^{-2}$) well developed (height 
$>$ 8 nm) QDs, there are many small surface features of minimum height of $\sim$ 14\AA 
$\,$and a mean height of 26\AA $\,$contributing to a large number of counts at low 
heights. Furthermore, in Fig. \ref{fig:fig2} (b)-(c), we observe that  these small surface 
features are  stable and continuously grow in size as more material is deposited. 
Therefore, 
\begin{figure}[!h]
\begin{center}
\resizebox{!}{10cm} {\includegraphics{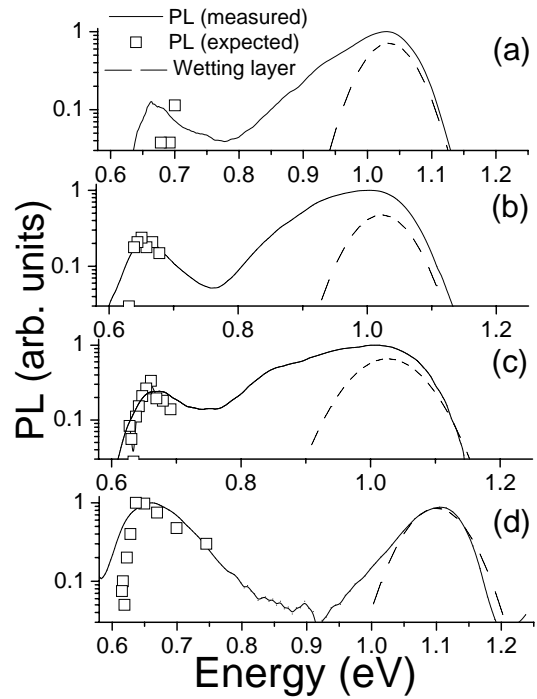}}
\caption{\label{fig:fig3}{\it (Solid line) 25K PL Spectra. The growth conditions for 
samples whose response is depicted in (a)-(d) are the same as their counterparts in Fig. 
\ref{fig:fig2}, except for the InP overgrowth,.  (Squares) Calculated transition energies 
\cite{holm-pryor} scaled with the heights histogram (Fig. \ref{fig:fig2}) in an attempt to 
reconstruct the PL spectra. The expected response from only the features higher than the 
cut-off depicted by dotted lines in Fig. \ref{fig:fig2} is shown. Note that there is no 
fitting parameter. The higher energy feature is fitted to two Gaussians. (Dotted line) 
Wetting layer contribution to PL. PL from the wetting layer appears at 1.02$-$1.03 eV in 
(a)-(c) and shifts to 1.10 eV for (d). See text.  The other fit corresponding to the 
evolution of small surface features is not shown here for brevity. }}
\end{center}
\end{figure}
with incremental coverage, Fig.\ref{fig:fig2}(b)-(d), the morphology evolves as (i) matter 
is accreted by all pre-existing features making them grow in size (ii) new well developed 
QDs of height $\sim$ 8nm are spontaneously generated (usual S-K transition). Growth via 
two independent pathways naturally leads to a bimodal size distribution at an intermediate 
stage of growth, Fig.\ref{fig:fig2}(c). The role of existing (two- and 
quasi-three-dimensional) surface features during the S-K transition has been much 
discussed\cite{Ramachandran, patella, priester}. Here these surface corrugations are only 
being connected to dots which continuously evolve in size and we have not examined their 
role, if any, as precursors to abruptly generated well developed dots (S-K transition).  

The above mentioned picture of growth evolution is independently observed in optical 
properties. Fig.\ref{fig:fig3} shows the low temperature PL spectra measured on the four 
similarly grown but capped samples. Along with the characteristic PL from well-developed 
dots, the PL spectrum also shows a wetting layer peak around 1.02$-$1.03 eV that is 
broadened towards lower energy by the emission from these optically active 
quasi-three-dimensional clusters. The spectrum in Fig.\ref{fig:fig3}(c) shows almost three 
distinct peaks corresponding to the wetting layer and the two kinds of quantum dots. 
Finally, in the saturation regime (when the number of well resolved QDs does not increase 
with coverage), the distribution of the dot sizes is again roughly unimodal, but with a 
large dispersion. 

The optical signature from the wetting layer was inferred by fitting two Gaussians, 
(corresponding to signals from WL and quasi-three dimensional clusters) to the high energy 
PL peak. We observe that the wetting layer peak is constant to within $10meV$ in 
Fig\ref{fig:fig3}(a)-(c). This low temperature PL peak at around 1.02$-$1.03eV corresponds 
to an approximately 4.5 ML strained InAs/InP quantum well\cite{leonelli}. This value may 
be used to calibrate the thickness with the deposition time, which is otherwise difficult 
to estimate with sub-ML precision in a typical MOVPE set-up without in-situ diagnostic 
tools. It is worth pointing out that there has been a wide variation in the reported value 
of the wetting layer thickness for InAs/InP, with the lower limit being $\sim$1ML 
\cite{girard, movpe refs3}. With a lower growth temperature, the wetting layer thickness 
is expected to be (exponentially) larger than its equilibrium value 
\cite{Johansson_and_seifert_kinetics1}, since the former is inversely dependent on the 
diffusion length \cite{snyder-mansfield-orr}. Because 450$^\circ$C is among the lower 
reported growth temperatures for InAs dots, a thicker wetting layer is naturally expected. 
Remarkably, a recent reflection high energy electron diffraction study\cite{gendry} on 
2D-3D transition in MBE grown InAs/InP films, reports for growth at 450$^\circ$C, a value 
of the wetting layer thickness that is very similar to what we have inferred from the PL 
spectra.

Finally, in Fig. \ref{fig:fig3}(d), the wetting layer PL shifts to a higher energy, 
$\sim$1.10 eV indicating the well known narrowing of wetting layer in the saturation 
regime \cite{wetting-layer-thinning}. 

The observed PL spectra from QDs should in principle be derivable from the quantitative 
analysis of the AFM images \cite{footnote_afm_pl_scaledifference}. Using published ground 
state energy calculations\cite{holm-pryor} for InAs/InP QDs, the expected emission spectra 
from {\em large} dots are shown in Fig.\ref{fig:fig3}(open squares).  While there is a 
good quantitative agreement without any fitting parameter for the expected and observed 
spectral features from large dots (those above the size cut-off depicted by dotted lines 
in Fig. \ref{fig:fig2}), the PL from the smaller features in each histogram did not agree 
as well and is therefore not shown in Fig.\ref{fig:fig3}. It has been previously 
established that the capping process can considerably change the morphology of the 
individual quantum dots and it is likely that the smaller dots are more affected by 
capping\cite{morphology change on overgrowth1, morphology change on overgrowth2} and 
compositional changes due to the As/P exchange during the early stage of InP overgrowth. 

\noindent
{\it Implications:} 
We have observed that the 2D-3D transition is not abrupt and the bimodality in the QD's 
size distribution at an intermediate stage of growth is a consequence of the growth 
evolving via two distinct pathways. These two observations are contrary to previous 
studies on InAs/GaAs\cite{petroff_prb, Ramachandran} and In$_x$Ga$_{1-x}$As/GaAs 
\cite{leon-fafard} 
where the 2D-3D transition had been found to be abrupt and thermodynamically first-order 
with the surface coverage playing the role of a critical parameter\cite{petroff_prb}. An 
abrupt appearance of quantum dots beyond a critical coverage has also been theoretically 
reproduced within a rate equation based model \cite{dobbs}. 

Bimodality in dot sizes also implies that the QDs ensemble cannot be characterized by a 
single length scale (e.g.,  mean island size)\cite{joyce review} during most of the 
evolution and rules out the very attractive possibility of data collapse onto a universal 
scaling\cite{ebiko} function during intermediate stages of growth. Absence of scaling at 
an intermediate growth stage has actually been observed by Krzyzewski, et al. (e.g., see 
ref. \cite{stangl}). The issue of bimodality in dot sizes at intermediate coverage, 
although not fully resolved, is typically explained by the presence of energy 
barriers\cite{bimodal equilibrium} corresponding to another (shape-change, from pyramid to 
dome) first order transition that has been experimentally studied in SiGe/Si \cite{sige 
science paper},  GaN/AlN \cite{GaN/AlN} and InAs/GaAs \cite{costantini-kern}. 
\begin{figure}[!h]
\begin{center}
\resizebox{!}{6cm} {\includegraphics{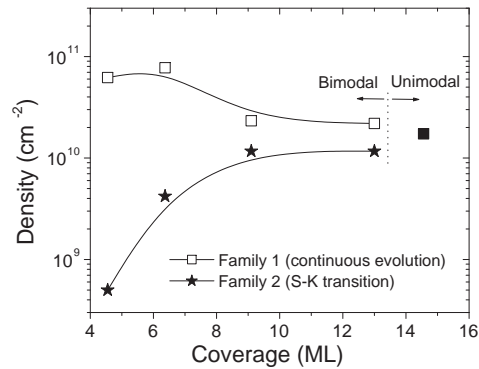}}
\caption{\label{fig:fig4}{\it Variation of the island density with coverage for the two 
families of islands corresponding to the two modes in the heights distribution in Fig. 1. 
Solid square represents the late stage when the distribution is umimodal (Fig. 
\ref{fig:fig2}(d)).}}
\end{center}
\end{figure}

It is instructive to (visually) separate \cite{da silva} the heights histograms in Fig. 
\ref{fig:fig2} at the minima between the two modes (dotted lines in Fig.\ref{fig:fig2}). 
Fig.\ref{fig:fig4} is the corresponding plot of cluster densities in these two families 
(S-K dots and continuously evolving clusters) as a function of coverage. The increase in 
the density of larger dots with coverage is abrupt, whereas the number of smaller features 
decreases with coverage due to coalescence. At large enough coverage the distribution 
becomes unimodal. Without invoking the existence of energy barriers, this is most simply 
understood as being due to the difference in growth rates of larger and smaller dots (due 
to their different volumes). 

Among the various studies of growth evolution, the results in references \cite{da silva, 
duarte}, although on MBE grown InAs/GaAs, are qualitatively most similar to ours. 
Nevertheless there are some important differences both in data and interpretation of 
results. Firstly, the quasi-three dimensional clusters were thought to be precursors to 
all the quantum dots giving a common origin to the dots in both the families.
Possibly due to a smaller strain in the InAs/InP system as compared to InAs/GaAs, we 
observe an uninhibited increase in size and a corresponding redshift in PL with coverage.  
This is in contrast with InAs/GaAs QDs, where a pronounced barrier seems to restrict the 
maximum dot size to $\sim$ 8nm. This supports the general observation of larger dispersion 
in dot sizes and low temperature PL linewidths ($\sim 100meV$) in InAs/InP. 

The differences observed in growth evolution by different groups may be a result of 
differing substrate/buffer layer conditions. It is possible for the local roughness in the 
wetting layer surface to stabilize pre-existing quasi-three dimensional surface structures 
and determine the extent of bimodality at an intermediate growth stage. Many of the 
thermodynamic arguments used to describe the QD self-assembly are relevant only under 
quasi-equilibrium conditions of growth which many such previous experiments tried to 
maintain. Our results, on the other hand, are obtained under much higher growth rates, 
more typical growth conditions of quantum dots. 

\noindent
{\em Conclusions}: Studying the growth evolution of MOVPE grown InAs/InP self-assembled 
quantum dots, we have observed that an alternate pathway for the 2D-3D transition exists 
that naturally explains the often observed bimodality in the quantum dots size 
distribution at an intermediate growth stage. Independent evaluations of the morphology 
and the photoluminesence spectra are consistent with the picture of growth presented.

We gratefully acknowledge J. John and Sandip Ghosh for their help with the AFM and PL 
measurements and Sandeep Krishna for his help with the development of the image processing 
software.\\  
}
\noindent
\small{$^*$Electronic Address: bhavtosh@tifr.res.in}


\begin{references}
\bibitem{footnote growth rate}This value of growth rate is inferred from the wetting layer 
PL in Fig.\ref{fig:fig3}(a).
\bibitem{footnote_afm_pl_scaledifference}It must be noted that AFM images cover an area of 
$1\mu m^2$ which is vastly different from the area sampled by the PL excitation source 
($\sim 1mm^2$). The comparison (which is further affected by InP overgrowth) is therefore 
necessarily qualitative.
\end{references}
\end{document}